\documentclass[a4paper]{jpconf}

\usepackage[english]{babel}
\usepackage{graphicx}
\usepackage{subfig}
\usepackage[frozencache,cachedir=minted-cache]{minted} 
\usepackage[hidelinks]{hyperref}
\usepackage{orcidlink}
\usepackage{xspace}
\usepackage{overpic}

\newcommand{\cpp}{C\nolinebreak\hspace{-.05em}\raisebox{.4ex}{\tiny\bf +}\nolinebreak\hspace{-.10em}\raisebox{.4ex}{\tiny\bf +}\xspace}

\begin{document}

\title{Challenges and opportunities integrating LLAMA into AdePT}

\date{2023-02-13}

\author{Bernhard Manfred Gruber\orcidlink{0000-0001-7848-1690}$^{1,3,4,5}$, Guilherme Amadio\orcidlink{0000-0002-2102-7945}$^1$ and\\Stephan Hageböck\orcidlink{0000-0001-9359-2196}$^2$}

\address{$^1$ EP-SFT, CERN, Geneva, Switzerland}
\address{$^2$ IT, CERN, Geneva, Switzerland}
\address{$^3$ Center for Advanced Systems Understanding (CASUS), Saxony, Germany}
\address{$^4$ Helmholz-Zentrum Dresden-Rossendorf (HZDR), Dresden, Germany}
\address{$^5$ Faculty of Computer Science, Technische Universität Dresden, Dresden, Germany}

\ead{bernhard.manfred.gruber@cern.ch}

\begin{abstract}
	Particle transport simulations are a cornerstone of high-energy physics (HEP),
	constituting a substantial part of the computing workload performed in HEP.
	To boost the simulation throughput and energy efficiency, GPUs as accelerators have been explored in recent years,
	further driven by the increasing use of GPUs on HPCs.
	The Accelerated demonstrator of electromagnetic Particle Transport (AdePT) is an advanced prototype for offloading the simulation of electromagnetic showers in Geant4 to GPUs, and still undergoes continuous development and optimization.
	Improving memory layout and data access is vital to use modern, massively parallel GPU hardware efficiently,
	contributing to the challenge of migrating traditional CPU based data structures to GPUs in AdePT.
	The low-level abstraction of memory access (LLAMA) is a \cpp library that provides a zero-runtime-overhead data structure abstraction layer, focusing on multidimensional arrays of nested, structured data.
	It provides a framework for defining and switching custom memory mappings at compile time to define data layouts and instrument data access, making LLAMA an ideal tool to tackle the memory-related optimization challenges in AdePT.
	Our contribution shares insights gained with LLAMA when instrumenting data access inside AdePT,
	complementing traditional GPU profiler outputs.
	We demonstrate traces of read/write counts to data structure elements as well as memory heatmaps.
	The acquired knowledge allowed for subsequent data layout optimizations.
\end{abstract}

\section{Introduction}

Particle transport simulations consume a substantial part of the computing resources in HEP:
11-24\% projected for 2031 by ATLAS~\cite{atlas_computing_roadmap},
15\% projected for 2031 by CMS~\cite{cms_offline_computing_report},
74.3\% used in 2021 and 75\% projected for 2024 by LHCb~\cite{lhcb_computing_usage_2021, lhcb_computing_2024_requests},
and around 70\% used in 2015 for ALICE\cite[p.30]{alice_upgrade_02_report}, summarized in~\cite{hep_requirements_talk}.
With the rise of accelerators, notably GPGPUs, new methods are in active development to boost simulation throughput and energy efficiency over classical and established simulation frameworks like Geant4~\cite{geant4_1, geant4_2, geant4_3}.
The Accelerated demonstrator of electromagnetic Particle Transport (AdePT) is a recent \cpp/CUDA prototype for offloading electromagnetic transport simulations to GPUs%
\footnote{AdePT is developed on GitHub: \url{https://github.com/apt-sim/AdePT}}
~\cite{adept_paper}.
It uses VecGeom~\cite{vecgeom} for handling geometry and G4HepEm~\cite{G4HepEm} for the electromagnetic physics implementation.
AdePT can run standalone, or via a fast simulation hook in Geant4.
Initial profiling of AdePT revealed that simulations are mostly bound by memory access and diverging code paths.
Independently, the low-level abstraction of memory access (LLAMA) has been developed as an answer to the continuously increasing memory-gap~\cite{llama_paper}:
Programs are increasingly memory-bound, so adapting the data layout for each target architecture is crucial.
LLAMA separates the algorithmic view of data from its mapping to device memory,
allowing different memory layouts to be chosen without touching the algorithm.
Given the capabilities of LLAMA, we concluded that the library provides an ideal toolbox to investigate AdePT's memory-related bottlenecks.
Furthermore, AdePT provides an ideal application to test LLAMA's capabilities for memory layout instrumentation, analysis, and optimization.
In this work, we present our findings when integrating LLAMA into AdePT.

\section{AdePT's track data structure}

\begin{figure}
	\centering
	\subfloat[Sparse single buffer]{
		\includegraphics[height=5cm, keepaspectratio, trim=0 0 0 15]{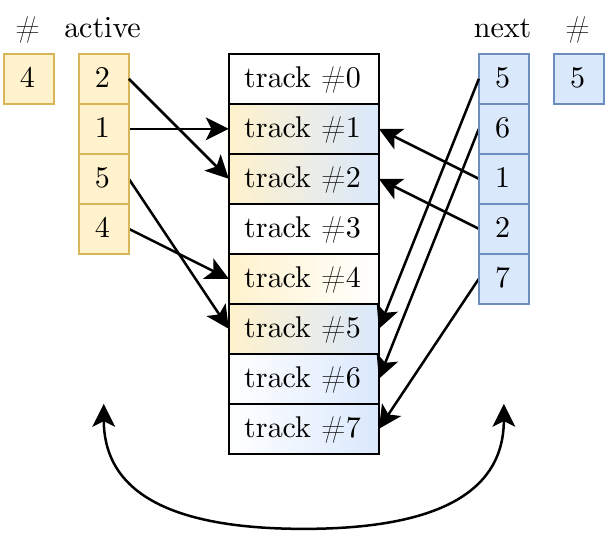}
		\label{fig:track_data_structure_sparse}
	}
	\subfloat[Dense double buffer]{
		\includegraphics[height=5cm, keepaspectratio, trim=0 0 0 15]{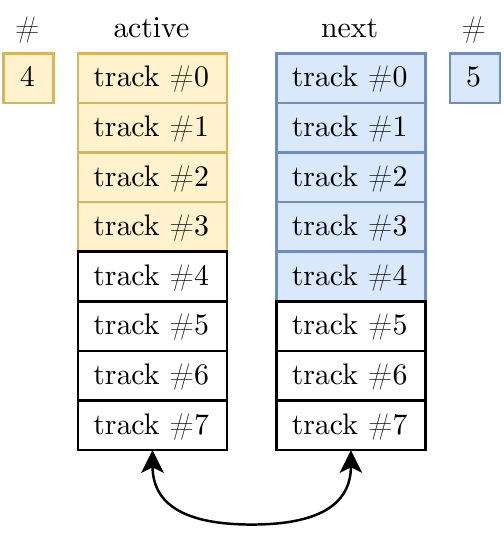}
		\label{fig:track_data_structure_dense}
	}
	\caption{
		The AdePT track data structure.
		Tracks are either saved linearly in an array using index queues (a)
        or can be kept dense using using two arrays without index queues (b).
	}
\end{figure}

The main data structure in AdePT is the list of active particle tracks.
By default, this is a sparse array of track structures, which is allocated once per particle type such as electrons, positrons and photons.
In addition to this track array, two lists of integers are required for managing the track slots of the current and subsequent transport iteration.
This version of the data structure is referred to as ``sparse single buffer'' throughout this paper and is visualized in figure~\ref{fig:track_data_structure_sparse}.

As part of this work, a new data structure has been developed.
Per particle type, it is comprised of two dense arrays of track structures,
one for the currently active particles and one for the particles active in the next iteration.
Separate lists of slots are thus obsolete.
This version of the data structure is called ``dense double buffer'' throughout this paper, and is visualized in figure~\ref{fig:track_data_structure_dense}.

\section{LLAMA integration}

For easy integration and convenience, LLAMA offers a single, amalgamated header file.
After integrating, the following code transformations were implemented, sketched in figure~\ref{fig:track_cpp_vs_llama}:
The AdePT track data structure, a \cpp struct, was formulated as a type list using the LLAMA record and field constructs.
Member functions of the track were converted to free functions, and functions with tracks as arguments or return values were converted to templates.
Instances and references to tracks and their data members were replaced with LLAMA constructs or deduced types (using \mintinline{cpp}{auto}).
Access to a data member of a track needed to be expressed as call with a tag type.
Pointers to CUDA memory became LLAMA views.
Finally, to increase accuracy in LLAMA's Trace mapping,
the existing CUDA kernels were refactored to tolerate the use of proxy references%
\footnote{A proxy-reference in \cpp is a user-defined type that acts like a language-built-in reference.}
to LLAMA-managed tracks.

\begin{figure}
	\centering
	\begin{minipage}{.49\textwidth}
		\begin{minted}[gobble=3,fontsize=\footnotesize]{cpp}

			struct Track {
			  // ...
			  Vector3D<Precision> pos;
			  NavStateIndex navState;


			  __device__ void InitAsSecondary(
			      const Track &parent) {


			    // ...
			    this->pos      = parent.pos;
			    this->navState = parent.navState;
			  }
			};
		\end{minted}
	\end{minipage}\hfill
	\begin{minipage}{.49\textwidth}
		\begin{minted}[gobble=3,fontsize=\footnotesize]{cpp}
			struct Pos {}; struct NavState {}; // ...
			using Track = llama::Record<
			  // ...
			  llama::Field<Pos, Vector3D<Precision>>,
			  llama::Field<NavState, NavStateIndex>>;

			template <typename SecondaryTrack>
			__device__ void InitAsSecondary(
			    SecondaryTrack &&track,
			    const Vector3D<Precision> &parentPos,
			    const NavStateIndex &parentNavState) {
			  // ...
			  track(Pos{})      = parentPos;
			  track(NavState{}) = parentNavState;
			}

		\end{minted}
	\end{minipage}
	\caption{
		The AdePT track data structure before and after the LLAMA integration.
		\label{fig:track_cpp_vs_llama}
	}
\end{figure}

\begin{figure}
	\centering
	\includegraphics[width=\textwidth]{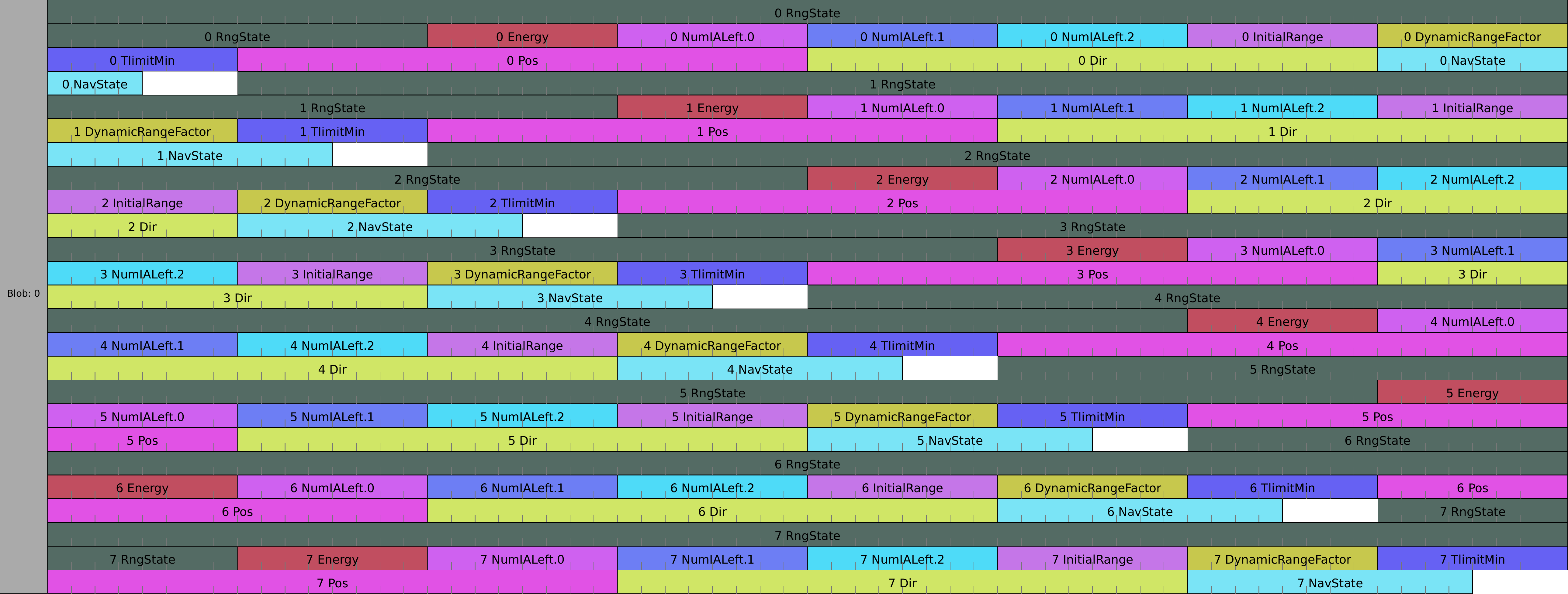}
	\caption{
		AdePT's default track layout, visualized by LLAMA.
		Memory is laid out left-to-right and wraps after 64 bytes.
		The boxes are colored by LLAMA field and labeled by array index and field name.
		\label{fig:tracklayout}
	}
\end{figure}

An immediate benefit of a data structure expressed with LLAMA is that LLAMA can visualize the data layout,
as shown in figure \ref{fig:tracklayout} for the default sparse single buffer.
The integration of LLAMA increased the compilation time for an incremental build of the benchmark executable by 27\% (one .cpp and three .cu files)
and required 178 insertions and 226 deletions on the 1336 lines of benchmark code
(as counted by the \texttt{cloc} utility, excluding external libraries).

\section{Memory layout benchmark}

\begin{figure}
\centering
	\centering
	\includegraphics[trim=25 8 20 10, width=0.96\textwidth]{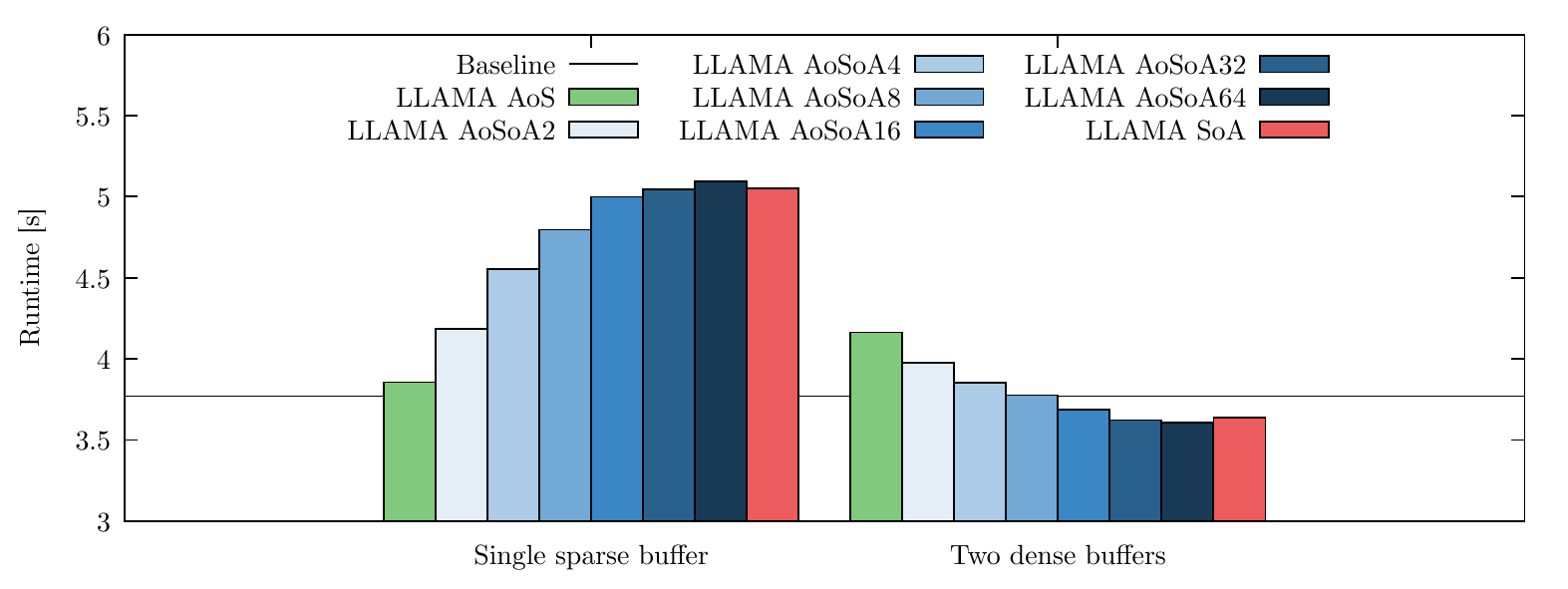}
	\caption{
		Runtime comparison between the single sparse and double dense buffer data structures using various LLAMA layouts.
		The reported numbers are the averages of 5 runs each.
		\label{fig:benchmark}
	}
\end{figure}

To benchmark the data structures and memory layouts we use AdePT's example19 and the TestEm3 geometry.
$10\,\mathrm{GeV}$ of highly energetic electrons are injected by a particle gun into 50 alternating box-shaped layers of absorber and gap material.
The simulation computes the development of an electromagnetic shower and the energy deposit in the detector's layers.
This test scenario provides a realistic compute workload, since it runs a complete physics implementation,
with a simple geometry, since there is ongoing R\&D to improve the geometry code in VecGeom.
All presented benchmarks%
\footnote{Benchmarks are available on GitHub: \url{https://github.com/apt-sim/AdePT_LLAMA_ACAT22}}
were compiled on a machine running CentOS Stream 8 with GCC 11 and CUDA 11.7 and run on an NVIDIA V100S%
\footnote{5120 CUDA cores, 80 SMs, 1597MHz clock speed, 8.2 TFLOPS in double precision, 32 GB HBM2/ECC memory and 1124 GB/S memory bandwidth}.
We used AdePT from GitHub (git commit 449222d), VecCore 0.8.0, VecGeom 1.1.20 and Geant4 11.1.0.
The following command line was used to run a benchmark:

\begin{minted}{bash}
$ example19 -particles 10000 -batch 5000 -gunpos -220,0,0 -gundir 1,0,0 \
            -gdml_file testEm3.gdml
\end{minted}

Figure \ref{fig:benchmark} shows the measured runtimes.
The baseline is the unmodified example19 (single sparse buffer).
Switching it to LLAMA AoS, which expresses the same data layout, incurs a small overhead of ~2.4\%.
Although LLAMA claims to have zero abstraction overhead, in our case the compiler failed to fully optimize all abstractions away, resulting in a different allocation of registers.
Progressing towards the SoA layout via multiple AoSoA versions using different lane counts%
\footnote{In an AoSoA$n$ layout, a record with the fields (A,B,C) and $n=2$ is laid out as AABBCCAABBCC etc.}
increases the runtime.
For the double dense buffer, the reverse effect was observed, with the AoS being the slower layout and runtime decreasing towards the SoA layout.
The fastest runtime was measured with the AoSoA64 layout.
It is worth noting that by using LLAMA the memory layout can be changed with a single line of code,
which allowed us fast and flexible experimentation with different memory layouts.

\section{Memory access instrumentation}

\begin{figure}
	\centering
	\includegraphics[width=0.96\textwidth, trim=25 5 20 10]{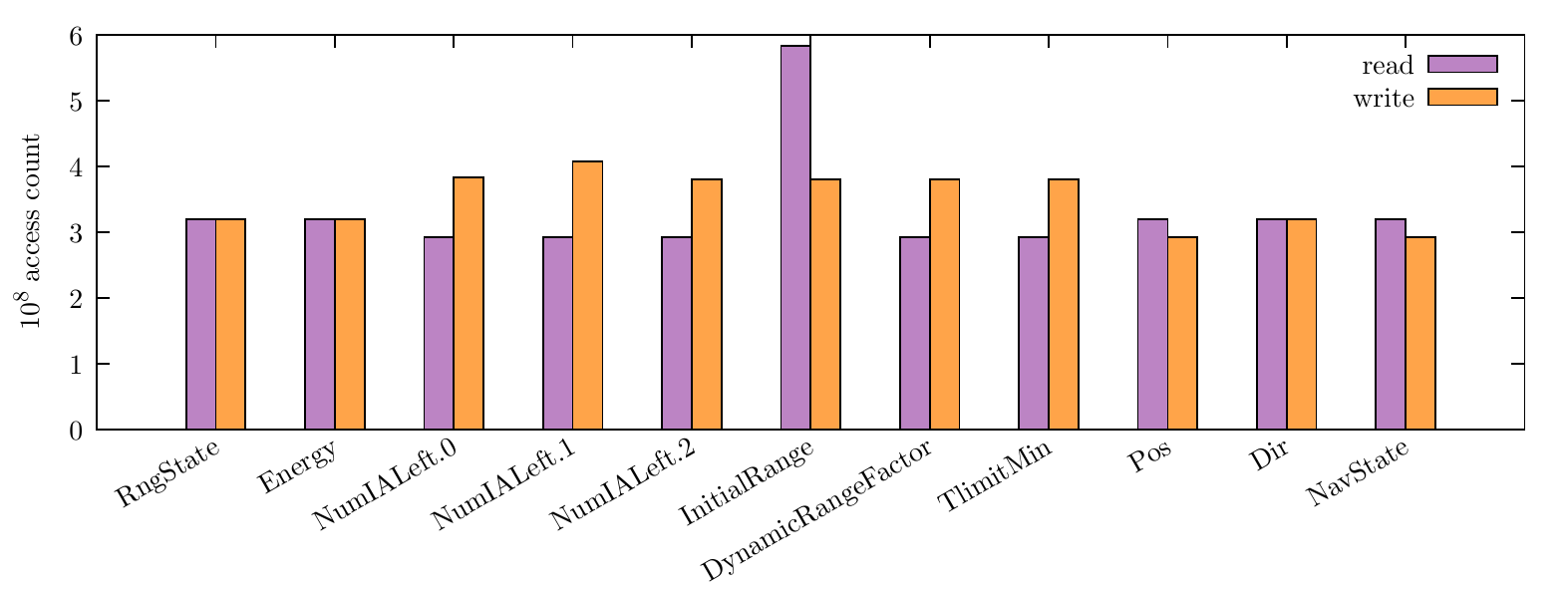}
	\caption{
		Access counts per track field on the electrons buffer as measured by the LLAMA Trace mapping.
		\label{fig:trace_electrons}
	}
\end{figure}

To better understand the benchmark results and the impact of memory layouts,
we used LLAMA's instrumentation mappings.
Detailed information is available in LLAMA's most recent publication\cite{llama_acat22_paper}.
Using LLAMA's lightweight Trace%
\footnote{The Trace mapping was renamed to FieldAccessCount in more recent versions of LLAMA.}
mapping we counted the accesses to each field of the electrons buffer.
These results are shown in figure \ref{fig:trace_electrons}.
Except for the InitialRange field, all other fields are almost uniformly read.
The write counts are less equally distributed and focus more on NumIALeft.0-2, InitialRange, DynamicRangeFactor and TlimitMin.
For those fields, the number of writes is also higher than the number of reads,
which is suspicious and requires further investigation.

\begin{figure}
	\centering
	\begin{overpic}[width=0.49\textwidth]{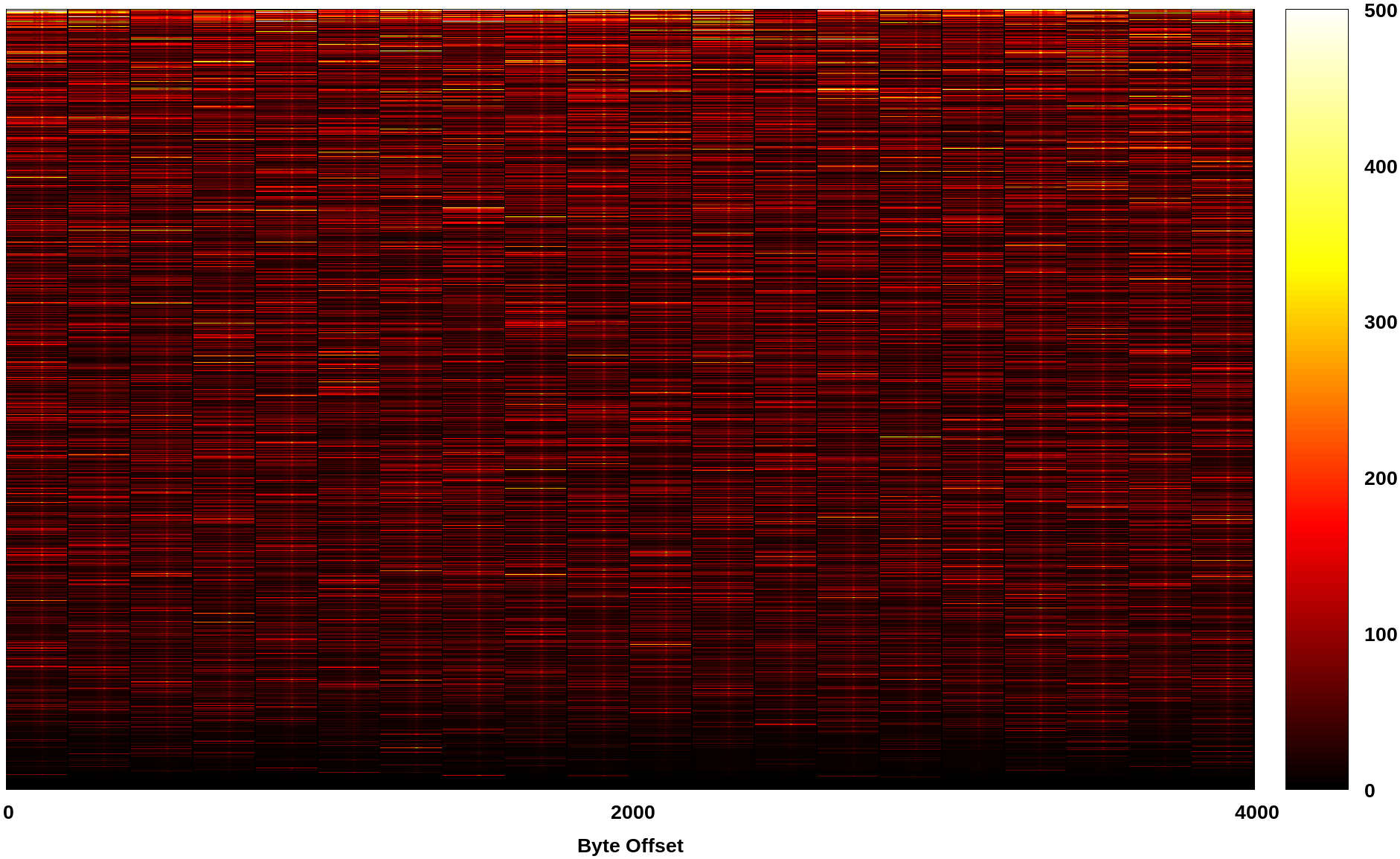}
		\put(3,7){\color{white}\textbf{Sparse AoS electrons}}
	\end{overpic}
	\begin{overpic}[width=0.49\textwidth]{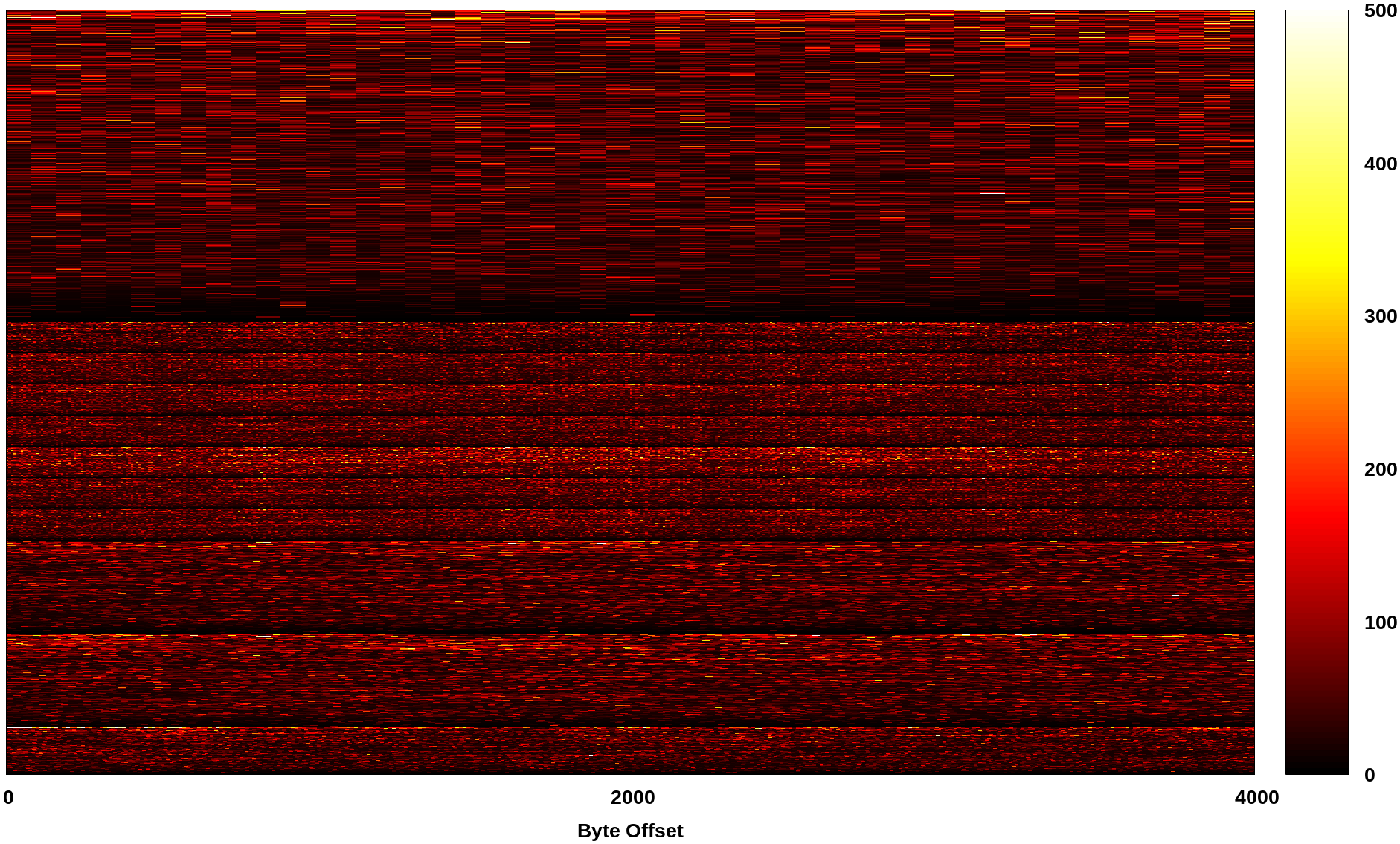}
		\put(3,7){\color{white}\textbf{Sparse SoA electrons}}
	\end{overpic}
	\\
	\begin{overpic}[width=0.49\textwidth]{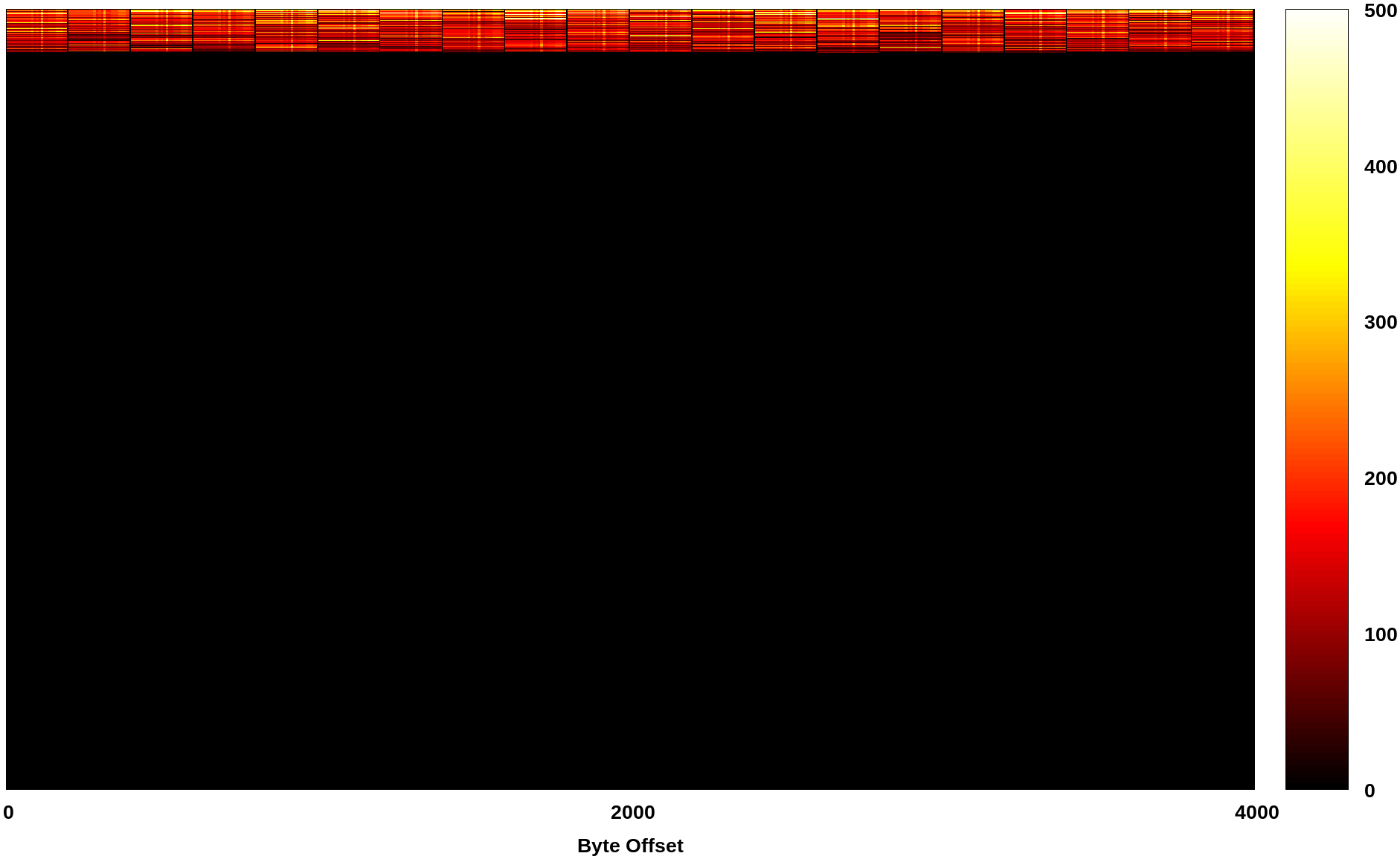}
		\put(3,7){\color{white}\textbf{Sparse AoS positrons}}
	\end{overpic}
	\begin{overpic}[width=0.49\textwidth]{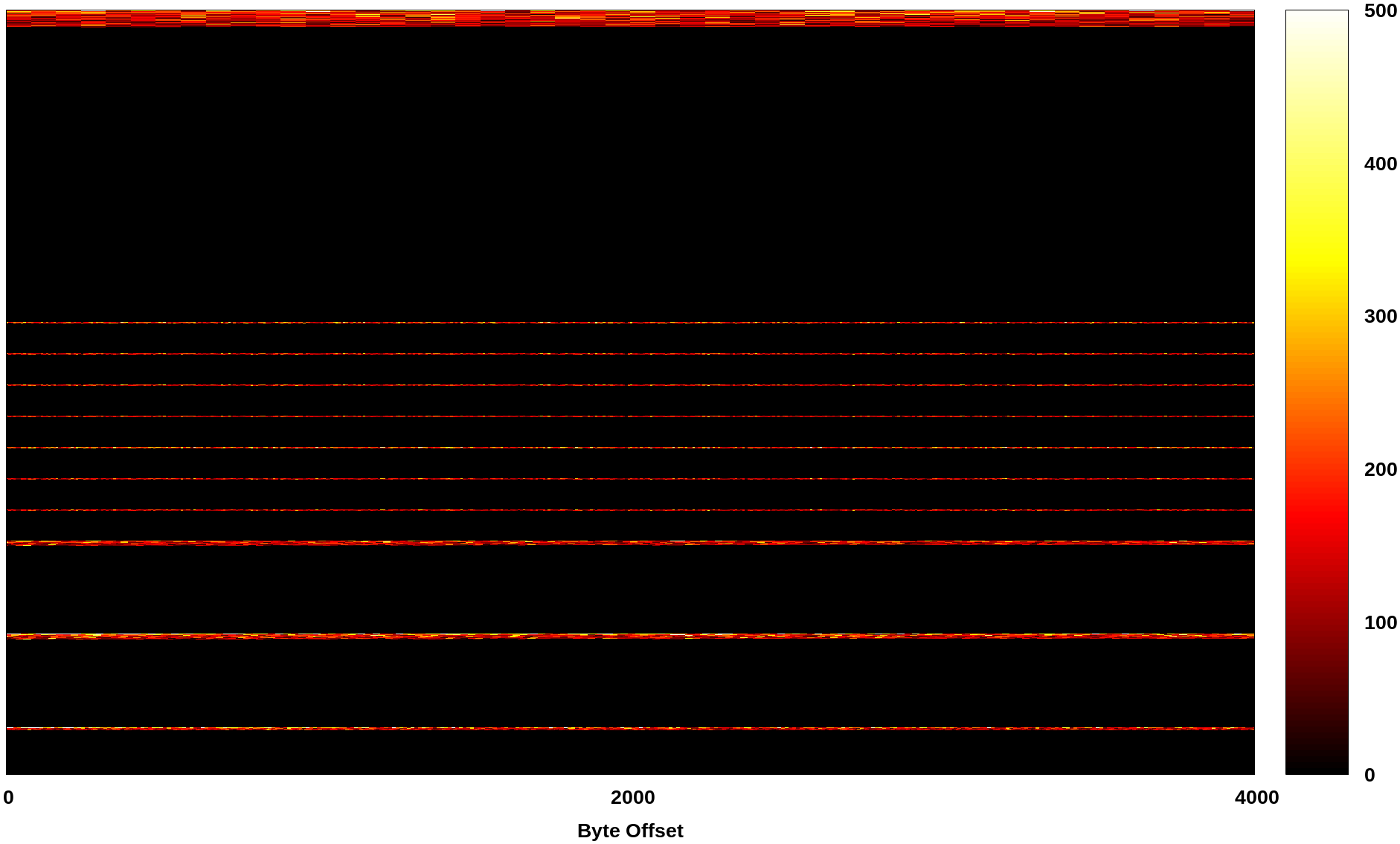}
		\put(3,7){\color{white}\textbf{Sparse SoA positrons}}
	\end{overpic}
	\\
	\begin{overpic}[width=0.49\textwidth]{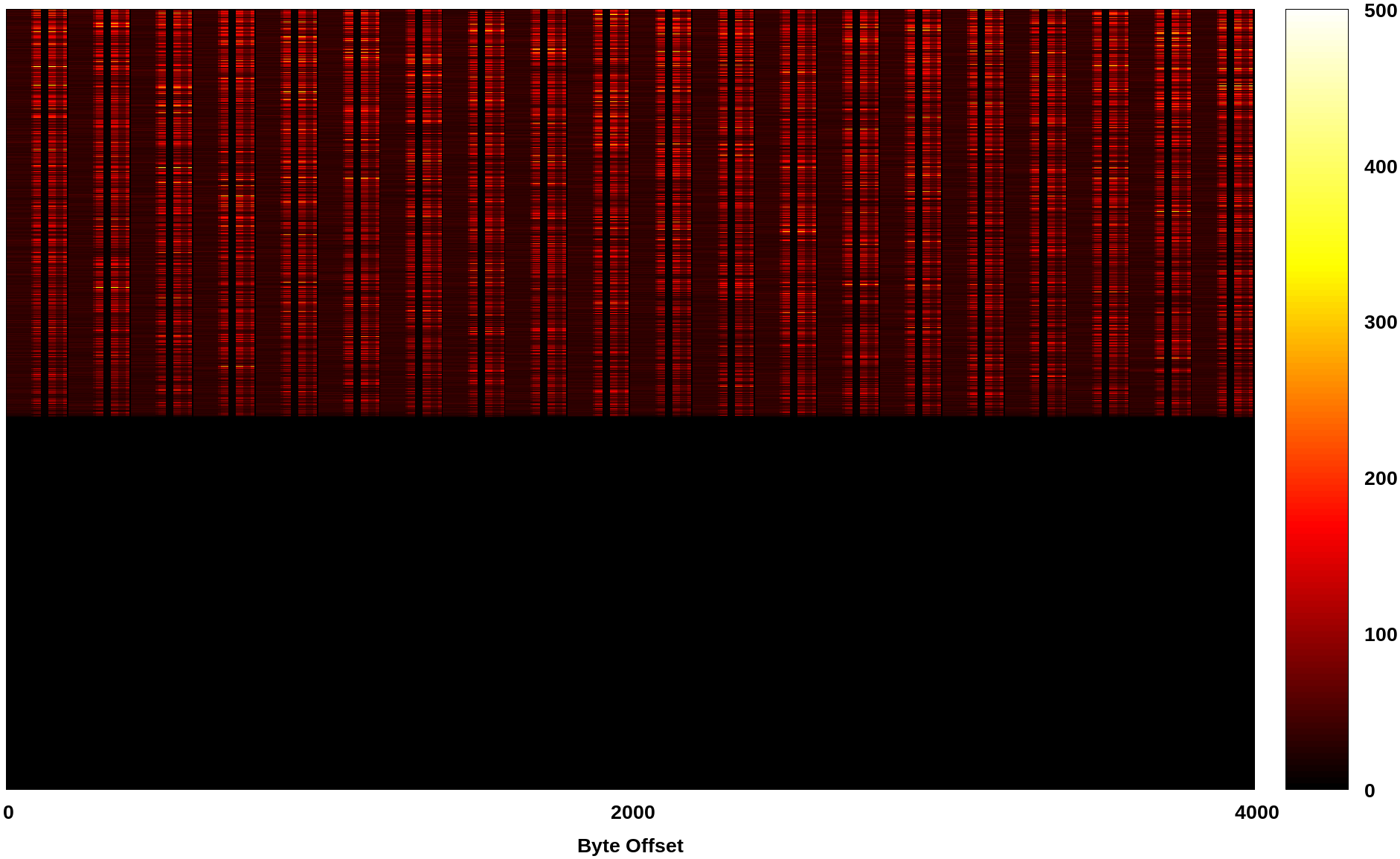}
		\put(3,7){\color{white}\textbf{Sparse AoS photons}}
	\end{overpic}
	\begin{overpic}[width=0.49\textwidth]{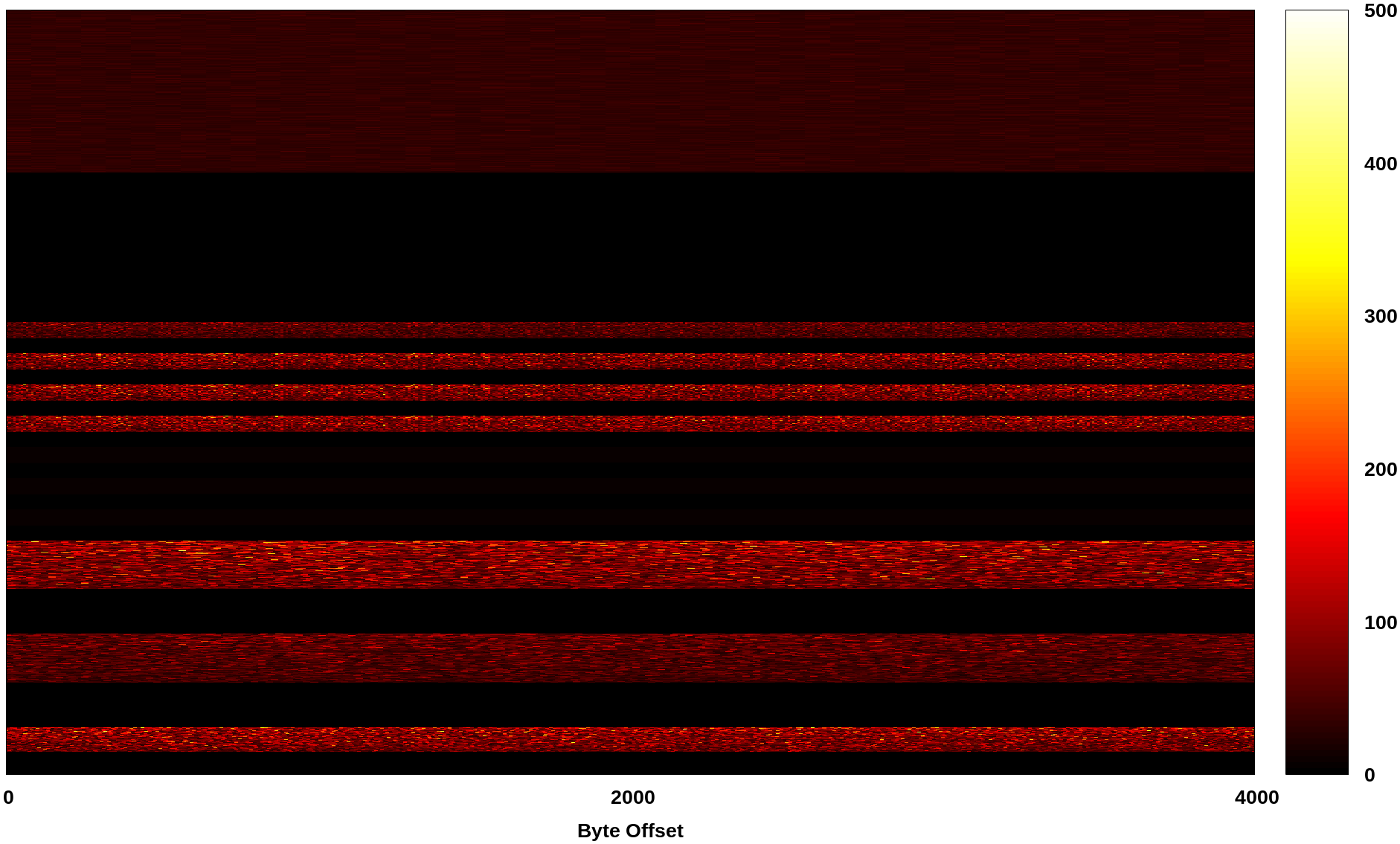}
		\put(3,7){\color{white}\textbf{Sparse SoA photons}}
	\end{overpic}
	\caption{
		LLAMA heatmaps for the single sparse buffer.
		One row shows 20 tracks.
		\label{fig:heatmaps_sparse}
	}
\end{figure}

Furthermore, LLAMA can create a heatmap counting the number of accesses per byte on various buffers, see figure~\ref{fig:heatmaps_sparse}.
Because this instrumentation has a high memory overhead,
we had to reduce the memory footprint of the simulation significantly
by only injecting 25 particles in 5 batches (\texttt{-particles 25 -batch 5}).
We observe, that the access is generally distributed rather randomly.
The electrons buffer is fully, the photons buffer halfway and the positrons buffer almost not utilized.

\begin{figure}
	\centering
	\begin{overpic}[width=0.32\textwidth,trim={49px 0 0 0},clip]{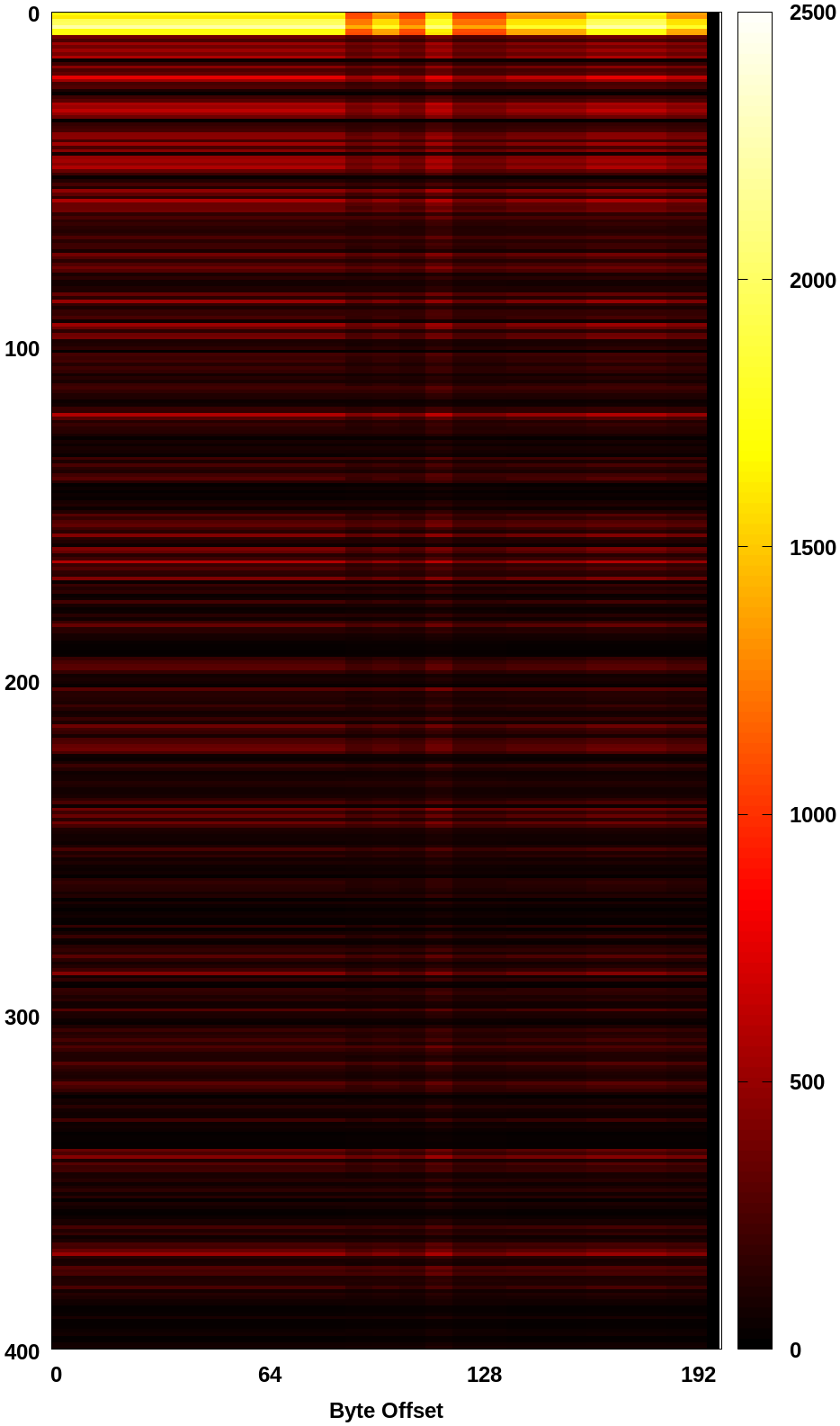}
		\put(3,7){\color{white}\textbf{Sparse electrons}}
	\end{overpic}
	\begin{overpic}[width=0.32\textwidth,trim={49px 0 0 0},clip]{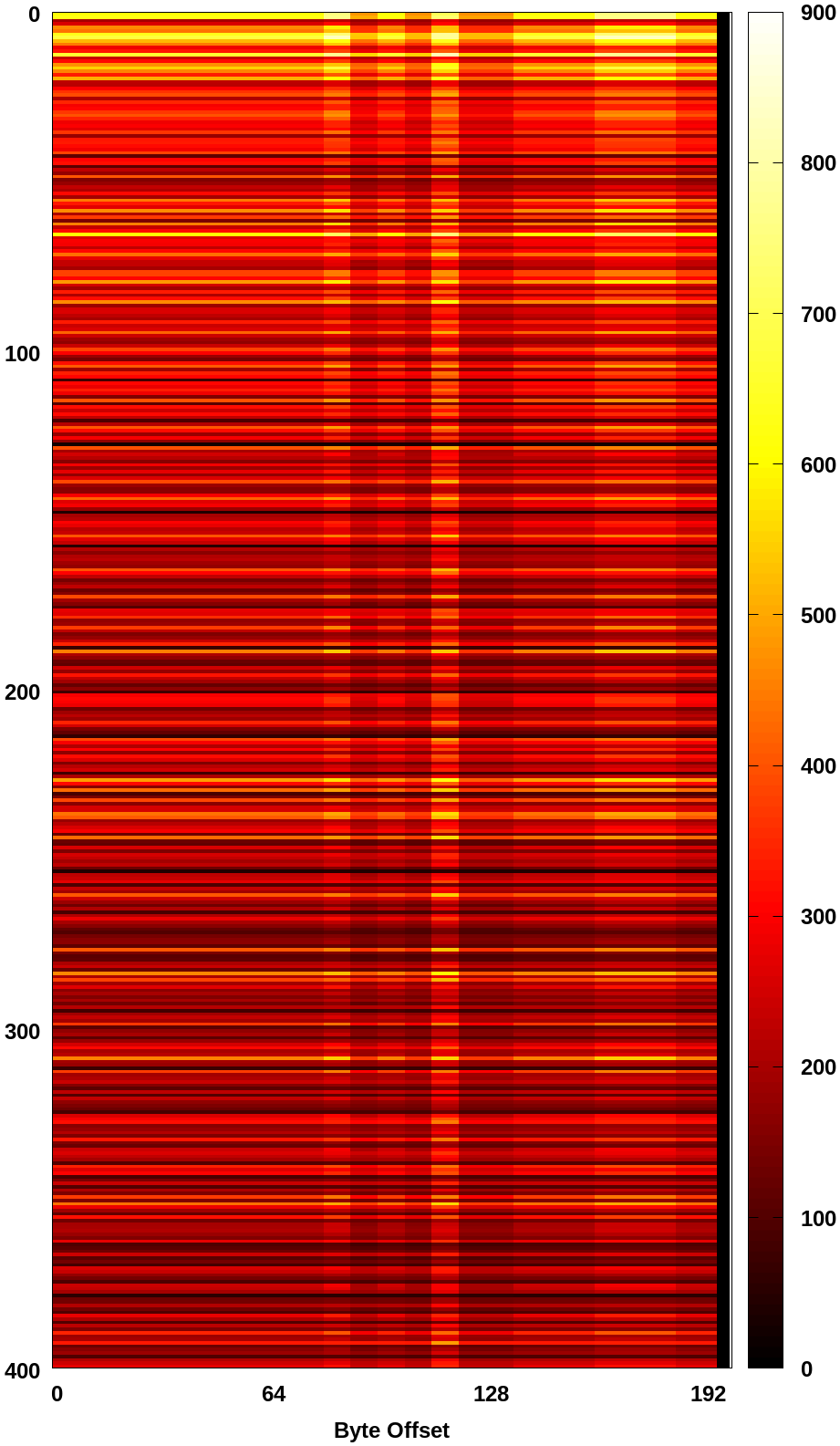}
		\put(3,7){\color{white}\textbf{Sparse positrons}}
	\end{overpic}
	\begin{overpic}[width=0.32\textwidth,trim={49px 0 0 0},clip]{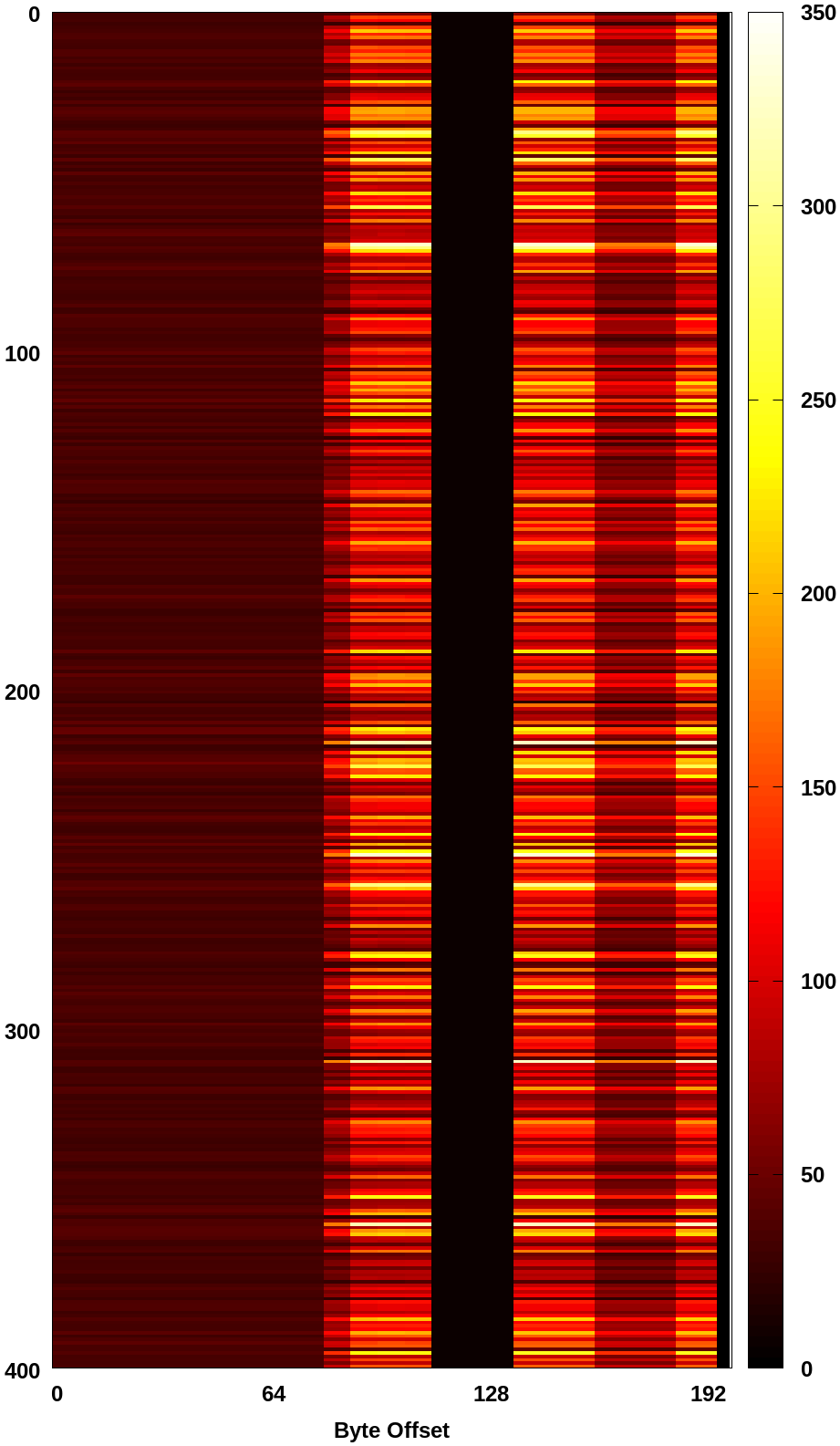}
		\put(3,7){\color{white}\textbf{Sparse photons}}
	\end{overpic}
	\caption{
		Magnified heatmaps for the single sparse buffer AoS layouts.
		One row is one track.
		\label{fig:heatmaps_closeup_sparse}
	}
\end{figure}

\begin{figure}
	\centering
	\begin{overpic}[width=0.32\textwidth,trim={49px 0 0 0},clip]{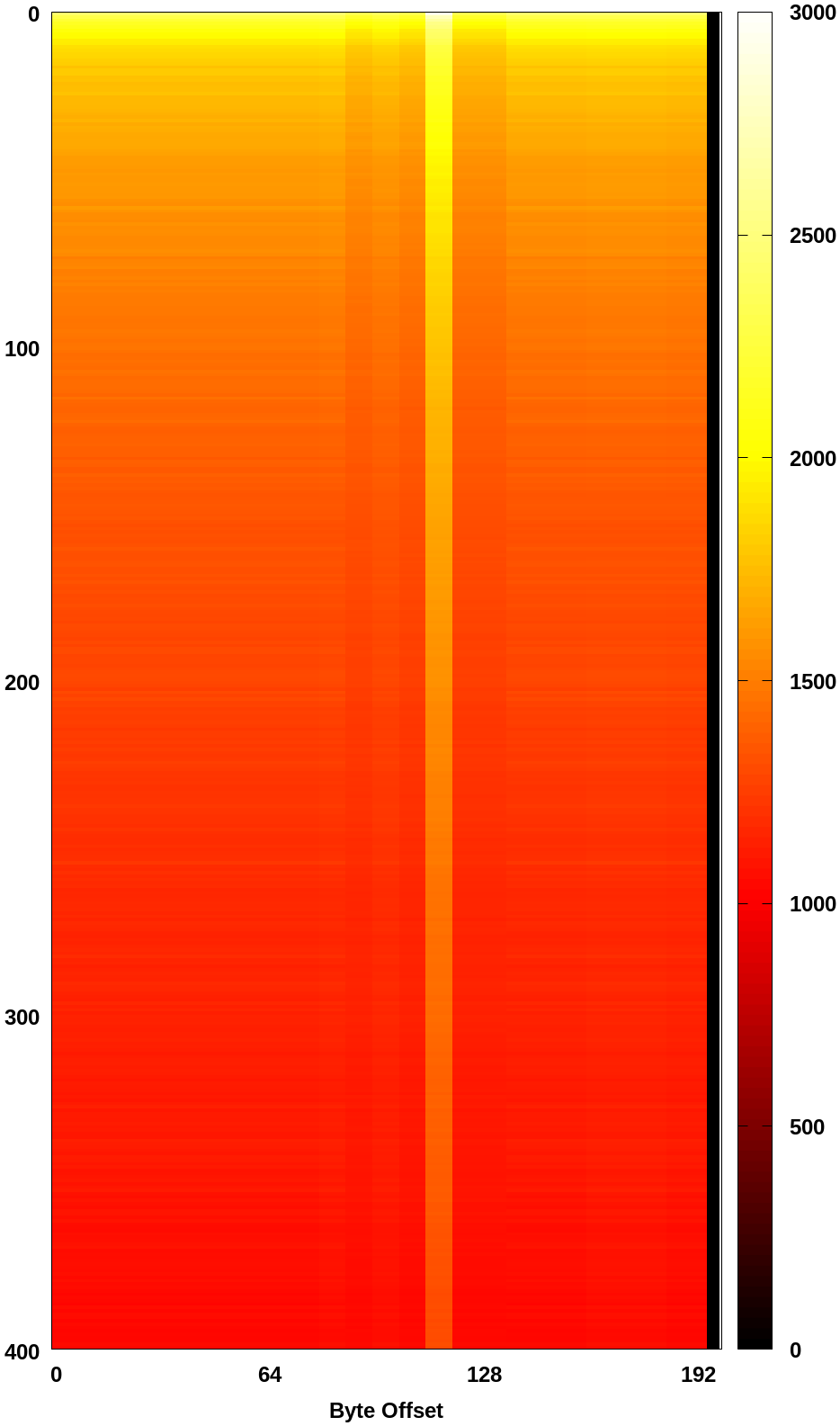}
		\put(3,7){\color{white}\textbf{Dense electrons}}
	\end{overpic}
	\begin{overpic}[width=0.32\textwidth,trim={49px 0 0 0},clip]{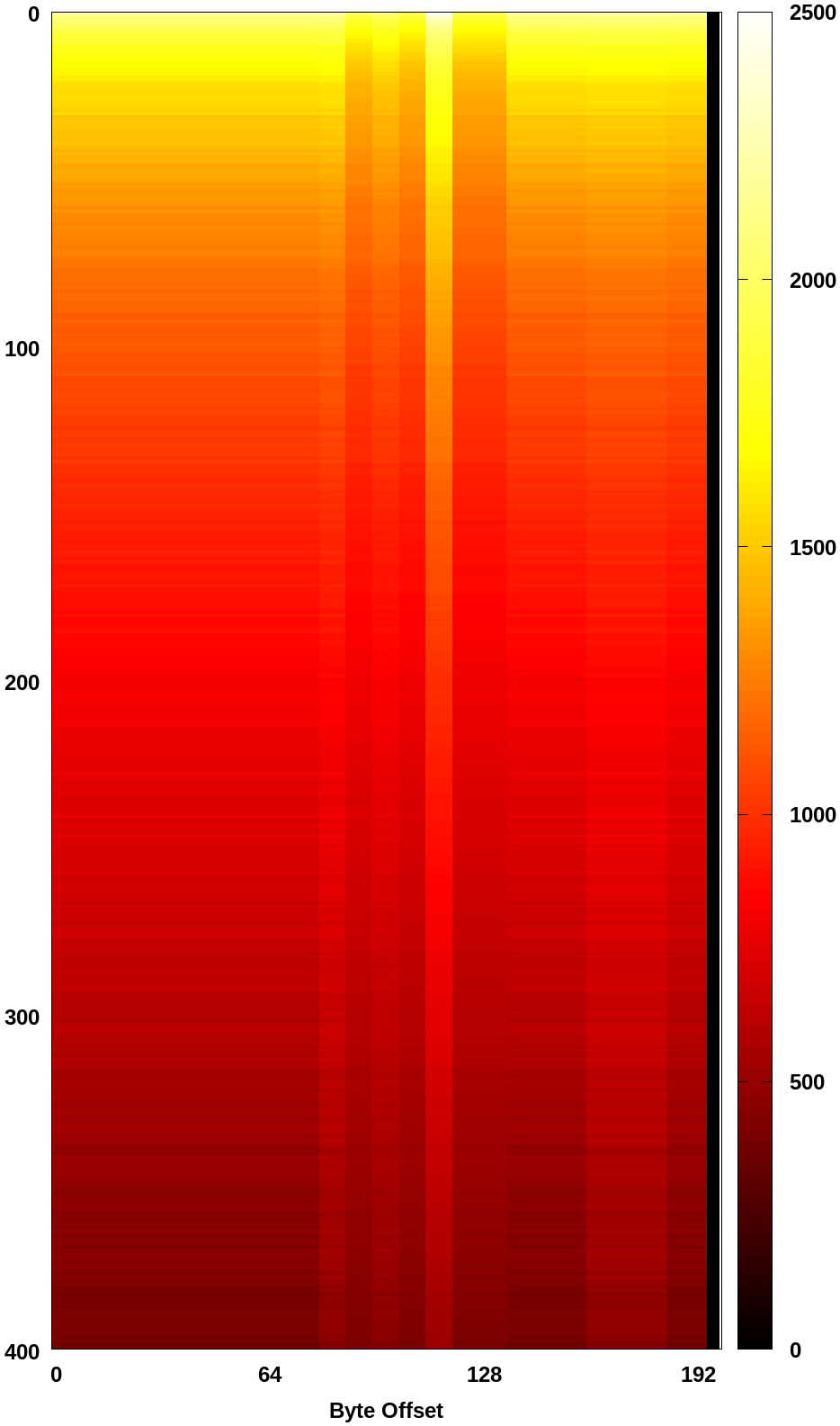}
		\put(3,7){\color{white}\textbf{Dense positrons}}
	\end{overpic}
	\begin{overpic}[width=0.32\textwidth,trim={49px 0 0 0},clip]{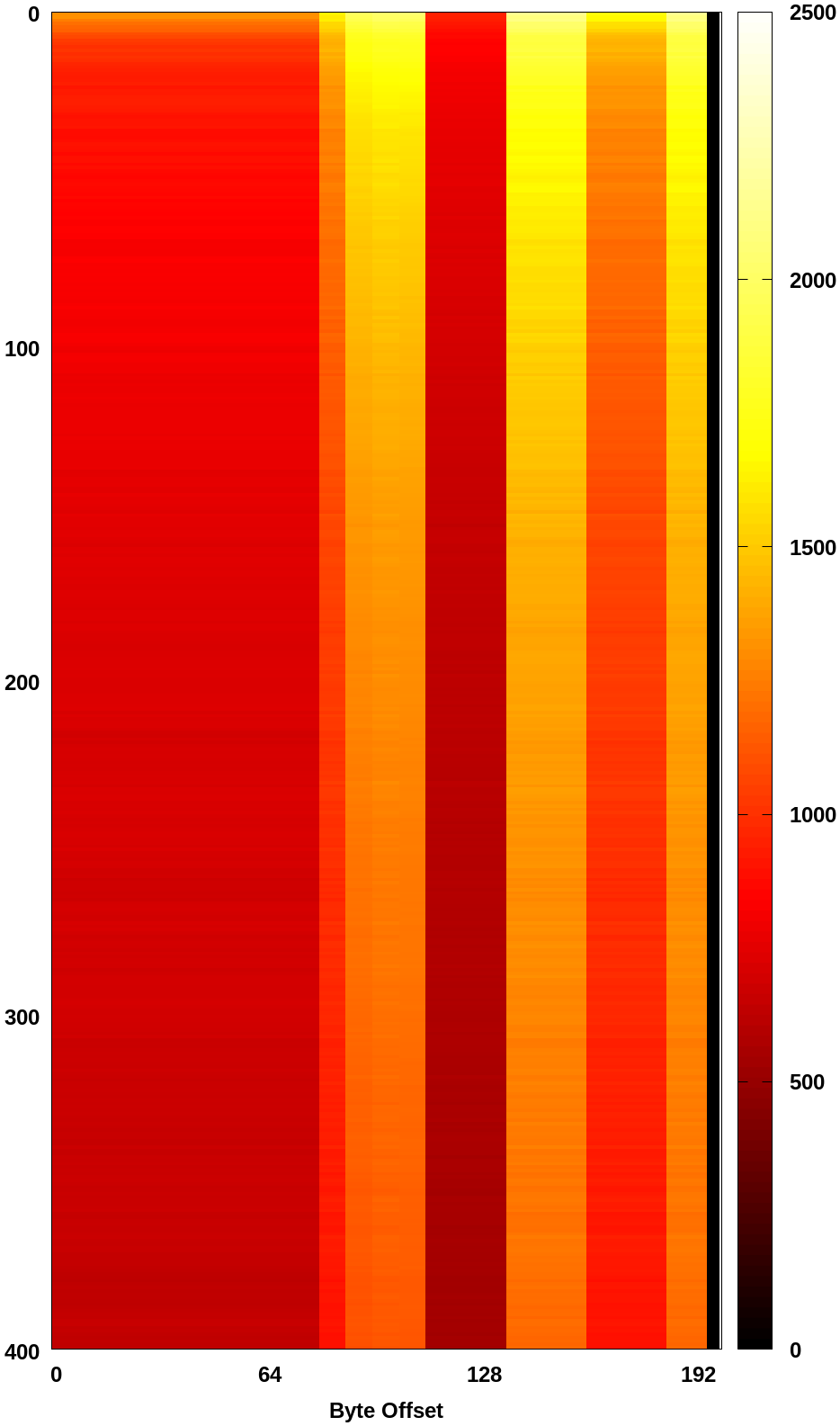}
		\put(3,7){\color{white}\textbf{Dense photons}}
	\end{overpic}
	\caption{
		Magnified heatmaps for the double dense buffer AoS layouts.
		One row is one track.
		\label{fig:heatmaps_closeup_dense}
	}
\end{figure}

Figure \ref{fig:heatmaps_closeup_sparse} shows a closeup of the AoS layouts for the electron, positron and photon sparse single buffers.
The heatmaps show that early-created electrons are accessed very frequently, unlike electrons produced later in the simulation.
A few electron fields are colder than the others and may be good candidates to split off to a different memory region.
Positrons are accessed more evenly across the array.
The photon random number generator state is accessed very infrequently in comparison to electrons and positrons.
This is confirmed by the implementation since photon physics requires far less random numbers.
The second-to-last field (direction) of electrons and positrons is hotter than for photons.
Some photon fields are barely accessed at all (InitialRange, DynamicRangeFactor, TlimitTime).
The maximum access count is significantly different,
with electrons being accessed more than 2.5 times as often as positrons and more than 7 times as often as photons.
There is tail padding in all buffers, which we have already seen in the memory layout visualization in figure \ref{fig:tracklayout}.
Figure \ref{fig:heatmaps_closeup_dense} shows a closeup of the AoS layouts for the electron, photon and positron double dense buffers.
Compared to the single sparse buffers in figure \ref{fig:heatmaps_closeup_sparse}, the double dense buffers exhibit a far more uniform access pattern.

The row sizes of the heatmaps can be chosen to study different properties.
Using the size of a single track emphasizes how the fields of a record are accessed.
Choosing a relevant hardware size, such as the cache-line size or memory bus width,
gives insight into how the data maps to hardware.

To inspect the usage of whole track slots in the buffers,
additional heatmaps were generated using an AoS layout with the size of a track as heatmap granularity.
For the single sparse electrons buffer, almost the entire buffer is required to provide track slots,
but the slots are barely used except for the first 5.
For the double dense buffers, track slot utilization is dominant at the front of the buffer,
keeping hot data much closer together.
As an added benefit, far less track slots are used in total (2044 vs.\ 45509 slots) allowing much bigger simulations to be run.

\section{Summary and conclusions}

We successfully integrated LLAMA into AdePT,
which allowed us to experiment with different memory layouts easily and fast.
With the AoSoA64 double dense buffer we were able to find a memory layout improving upon the status quo.
We showed that the SoA layout is not a silver bullet and requires a dense access pattern --
memory layout and access pattern must fit together.
The AoS layout works substantially better with sparse and random access.
AoSoA layouts with various blocking factors balance between AoS and SoA.
All of these could be tested in an afternoon after LLAMA was integrated into AdePT.

LLAMA's memory access visualization gave us useful insights for future research.
Even though all particle types use the same data structure, the different access patterns may warrant different memory layouts.
Individualizing, splitting and regrouping parts of the data structure could improve AdePT's performance.
A separation of hot and cold data would be a first step.
The heatmaps and data layout visualizations also proved useful for studying
how the data structure design affects padding, coalescing and mapping to cache lines.

The compile time increase and small overhead caused by LLAMA's abstractions
is greatly offset by the gained flexibility and instrumentation capabilities,
although the necessary code changes are invasive.
With LLAMA integrated, we have great tools to our aid for subsequent performance optimization.

\ack

This work has been sponsored by the Wolfgang Gentner Programme of the German Federal Ministry of Education and Research (grant no. 13E18CHA).
The primary author would like to thank Verena Gruber for proof-reading and commentary.

\section*{References}
\bibliography{bibliography}

\end{document}